\documentclass[letterpaper,twocolumn,10pt]{article}
\usepackage{usenix2019_v3}

\usepackage{listings}
\usepackage{amsmath}
\usepackage{amssymb}
\usepackage{xcolor} 
\usepackage{xspace}
\usepackage{soul}

\usepackage{tikz}
\usepackage{amsmath}
\usepackage{amsmath}

\usepackage{threeparttable}
\usepackage{xcolor}
\usepackage{multirow}
\usepackage{booktabs}
\usepackage{graphics}
\usepackage{siunitx}
\usepackage{soul}

\usepackage{graphicx}    
\usepackage{tcolorbox}   
\usepackage{caption}     

\usepackage{tabularx}
\newcolumntype{s}{>{\hsize=.5\hsize}X}
\newcolumntype{C}{>{\centering\arraybackslash}X}

\newcommand{\system}{LLMxCPG\xspace}
\newcommand{\systemquery}{LLMxCPG-Q\xspace}
\newcommand{\systemclassifier}{LLMxCPG-D\xspace}

\newcommand\khang[1]{{\color{black}#1}}

\newcommand\revision[1]{{\color{black}#1}}

\begin{document}

\date{}

\title{LLMxCPG: Context-Aware Vulnerability Detection Through Code Property Graph-Guided Large Language Models}

\author{
\emph{Ahmed Lekssays\textsuperscript{1*}, Hamza Mouhcine\textsuperscript{1*}, Khang Tran\textsuperscript{2}, Ting Yu\textsuperscript{3}, Issa Khalil\textsuperscript{1}}\\
\textsuperscript{1}Qatar Computing Research Institute, \textsuperscript{2}New Jersey Institute of Technology, \\
\textsuperscript{3}Mohamed bin Zayed University of Artificial Intelligence\\
\texttt{\{alekssays, hmouhcine, ikhalil\}@hbku.edu.qa}, \\\texttt{kt36@njit.edu}, \texttt{ting.yu@mbzuai.ac.ae} \\
\textit{* Joint first authors with equal contribution}
}

\maketitle

\thispagestyle{empty}
\pagestyle{empty}

\begin{abstract}
Software vulnerabilities present a persistent security challenge, with over 25,000 new vulnerabilities reported in the Common Vulnerabilities and Exposures (CVE) database in 2024 alone. While deep learning based approaches show promise for vulnerability detection, recent studies reveal critical limitations in terms of accuracy and robustness: accuracy drops by up to 45\% on rigorously verified datasets, and performance degrades significantly under simple code modifications. This paper presents \system, a novel framework integrating Code Property Graphs (CPG) with Large Language Models (LLM) for robust vulnerability detection. Our CPG-based slice construction technique reduces code size by 67.84 to 90.93\% while preserving vulnerability-relevant context. Our approach's ability to provide a more concise and accurate representation of code snippets enables the analysis of larger code segments, including entire projects. This concise representation is a key factor behind the improved detection capabilities of our method, as it can now identify vulnerabilities that span multiple functions. Empirical evaluation demonstrates \system's effectiveness across verified datasets, achieving 15-40\% improvements in F1-score over state-of-the-art baselines. Moreover, \system maintains high performance across function-level and multi-function codebases while exhibiting robust detection efficacy under various syntactic code modifications.
\end{abstract}

\section{Introduction}
\label{sec:introduction}

Software vulnerabilities continue to pose significant security risks, with the Common Vulnerabilities and Exposures (CVE) database reporting over 25,000 new vulnerabilities in 2024 alone \cite{statista_cve_2024}. Detecting these vulnerabilities early in the development life cycle is crucial for preventing security breaches and maintaining software integrity. However, despite extensive research, identifying vulnerabilities in complex codebases remains a challenging problem.

Recent approaches leveraging deep learning models have shown promise in vulnerability detection \cite{shimmi2024vulsim, chakraborty2021deep, chen2023diversevul, lu2021codexglue}. However, these approaches face several critical limitations that hinder their practical application. First, they typically focus on function-level analysis, overlooking crucial inter-procedural dependencies and broader program context \cite{primevul, ivdetect, russell, vuldeepecker, devign, sysevr, reveal}. Second, recent comprehensive evaluations have exposed weaknesses in their reliability. In a thorough assessment, Ding et al. \cite{primevul} introduced PrimeVul, a rigorously verified dataset where vulnerability labels were validated through multiple rounds of expert review and dynamic analysis. When state-of-the-art models were evaluated on this dataset, they exhibited dramatic performance degradation, with accuracy dropping by up to 45\% compared to their reported results on traditional datasets. This significant performance gap suggests that these models may be learning superficial patterns rather than meaningful vulnerability indicators. This hypothesis is further supported by Risse et al. \cite{bohme_limits}, who demonstrated significant performance drops when evaluating these models on datasets with simple modifications such as changes to function or parameter names. Finally, many approaches \cite{shimmi2024vulsim, hanif2022vulberta} are constrained by small embedding models with limited context windows, restricting their ability to analyze large code segments effectively.

To address these fundamental challenges, we present \system, a novel approach that combines Code Property Graphs (CPG) with Large Language Models (LLM) for robust vulnerability detection. Our approach systematically addresses the limitations of existing methods through its technical architecture. First, to overcome the context limitation and enable effective analysis of large codebases, \system introduces a sophisticated CPG-based analysis that converts input code into precise vulnerability-focused code slices. This slice construction process works in three phases: i) extracting potential vulnerable execution paths using the Static Application Security Testing (SAST) tool Joern and its CPGQL query language, ii) analyzing execution paths through CPG traversal to identify code elements that interact with the execution paths, and iii) constructing focused code snippets containing only essential components related to potential vulnerabilities by applying backward slicing. Second, to ensure robust feature learning beyond superficial patterns, we leverage these focused code slices to fine-tune a large language model specifically for vulnerability detection, enabling it to learn from and identify vulnerability patterns in concise and relevant code contexts.

Our empirical analysis demonstrates how \system successfully addresses the limitations of existing approaches. The slice construction approach achieves significant code reduction ratios, ranging from 67.84\% for function-level datasets to 90.93\% for multi-function codebases. By focusing the model's attention on these concise, vulnerability-relevant code segments rather than entire codebases, \system enables more effective learning of vulnerability characteristics. This targeted learning translates directly into robust performance: unlike existing approaches that show significant degradation on high-quality datasets, \system maintains consistent performance across both traditional datasets and rigorously verified ones. The effectiveness of our focused learning approach is reflected in substantial improvements over state-of-the-art baselines, with increases of 15\%-40\% in F1-score and 9-27\% in Accuracy on function-level real-world vulnerability detection tasks. Moreover, while current approaches struggle with complex codebases, \system demonstrates strong generalization capabilities on both function-level and multi-function codebases, performing effectively in scenarios where existing approaches fail to exceed random-guess performance. The system's robustness is further evidenced by its consistent detection efficacy under various syntactic modifications while preserving semantic equivalence, directly addressing the brittleness observed in current approaches.

\textbf{Contributions.} The contributions of this paper are summarized as follows:
\begin{itemize}
    \item \textit{Integration of LLMs with Program Analysis.} We present a novel framework that effectively combines traditional program analysis techniques (CPG) with modern large language models, creating a hybrid approach that leverages the strengths of both methodologies for improved vulnerability detection.
    \item \textit{Vulnerability-Focused Slice Construction.} We introduce a sophisticated code slicing technique that leverages Code Property Graphs (CPG) to extract vulnerability-relevant code segments that capture the essential elements of potential vulnerabilities while eliminating irrelevant code.
    \item \textit{Generalizability to Complex Codebases.} We demonstrate the generalizability of \system to both unseen function-level datasets and real-world open-source software.
    \item \textit{Robustness against Code Transformations.} We evaluate \system under different code transformations and show that it maintains its detection efficacy. 
    \item \textit{Open-Source Code and Datasets.} All source code and datasets used in this study are open-source, supporting reproducibility, transparency, and further research in the domain of software security.
\end{itemize}

\textbf{Outline.} The remainder of this paper is organized as follows: Section \ref{sec:background} provides background on vulnerability detection and code property graphs. Section \ref{sec:methodology}  details our \system approach. Section \ref{sec:evaluation} presents our experimental setup, evaluation methodology, and results. Section \ref{sec:discussion} discusses our results and their implications. Section \ref{sec:related_work} reviews related work, and Section \ref{sec:conclusion} concludes with future research directions.

\section{Background}
\label{sec:background}

\subsection{Vulnerability Detection}
\label{subsec:background_vuln_detection}

Software vulnerability detection remains a critical research area in computer security, with approaches broadly categorized into static and dynamic analysis techniques. On the one hand, dynamic analysis techniques observe program behavior during execution through methods such as fuzzing, and dynamic taint tracking \cite{liu2024source}. While these approaches provide more precise vulnerability detection by analyzing actual program execution paths, they face challenges in achieving comprehensive code coverage and handling the exponential growth of execution paths. Modern approaches increasingly leverage large language models and hybrid techniques that combine static and dynamic analysis, demonstrating improved detection capabilities while managing computational overhead \cite{primevul}.

 Static analysis methods, on the other hand, examine program code without execution, utilizing techniques such as pattern matching, data flow analysis, and abstract interpretation to identify potential security flaws \cite{steenhoek2023empirical}. These approaches offer comprehensive coverage but often generate false positives due to the inability to verify runtime behavior. Recent advancements in machine learning-based vulnerability detection have shown promise in improving detection accuracy, though challenges remain in handling complex codebases and reducing false positives \cite{chen2023diversevul}. In this work, we focus on developing novel static code analysis tools powered by LLMs to enhance vulnerability detection performance on real-world code snippets.

\subsection{Code Property Graphs}
\label{subsec:cpg}

Code Property Graphs (CPGs) represent a unified approach to program analysis by merging multiple code representations into a single graph structure \cite{yamaguchi2014modeling}. This representation combines abstract syntax trees (ASTs), control flow graphs (CFGs), and program dependence graphs (PDGs) into a joint data structure, enabling comprehensive analysis of code properties. The CPG preserves the syntactic structure from ASTs, control flow information from CFGs, and both control and data dependencies from PDGs, allowing complex patterns to be expressed through graph traversals. Analysts can express patterns through graph traversals using custom graph  query languages.
These traversals can be efficiently executed using graph databases, making it practical to analyze large-scale software projects. 


Due to this advantage, CPGs have emerged as a particularly effective tool for vulnerability detection. By combining syntactic, control flow, and data dependency information in a unified representation, CPGs enable the precise formulation of vulnerability patterns through graph traversals. This comprehensive view allows security analysts to express complex vulnerability patterns that would be difficult to capture using traditional static analysis approaches. For instance, CPGs can effectively model patterns for buffer overflows by simultaneously analyzing allocation operations in the abstract syntax tree, validating control flow paths for proper bound checking, and tracking data dependencies to identify attacker-controlled input \cite{liu2024source}. In this work, we employ Joern \cite{yamaguchi2014modeling}, an open-source static analysis tool that generates CPGs and supports queries over CPGs using the CPGQL language. 

\section{Methodology}
\label{sec:methodology}

\subsection{System Overview}
\label{subsec:system_overview}

\begin{figure*}[h]
    \centering
    \includegraphics[scale=0.11]{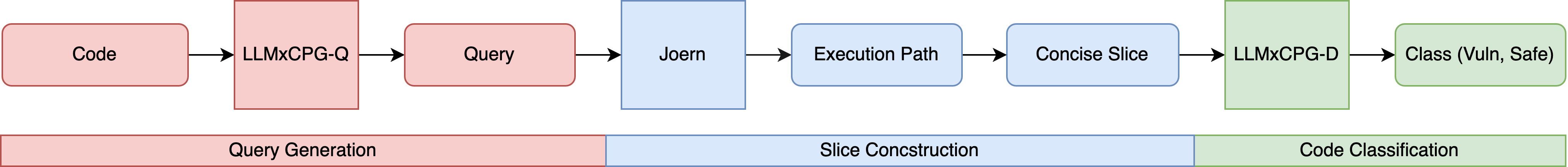}
    \caption{System Overview}
    \label{fig:system_overview}
\end{figure*}

We propose \system, a novel approach that capitalizes on the structural representation provided by CPGs while leveraging the sophisticated pattern recognition and generative abilities of Large Language Models (LLMs) for code vulnerability detection as depicted in Figure~\ref{fig:system_overview}.
Our approach comprises two specialized models, each fine-tuned for distinct tasks in our two-phase process. The first model, \systemquery, focuses on the slice construction phase by generating CPGQL queries that identify potentially vulnerable execution paths within the code. These queries enable the extraction of focused, security-critical code segments. The second model, \systemclassifier, handles the classification phase by analyzing these extracted code slices to determine their vulnerability status, categorizing them as either Vulnerable or Safe. 
This dual-model architecture combines the precision of graph-based program analysis with the advanced reasoning capabilities of state-of-the-art language models, enabling more accurate and interpretable vulnerability detection compared to traditional approaches. We note that the two models \systemquery~and \systemclassifier~are finetuned from Qwen2.5-Coder-32B Instruct and QwQ-32B-Preview, respectively (see Section \ref{subsec:implementation} for more details).
We note that, in what follows, when we refer to \system, we are referencing the entire process, including both slice construction and vulnerability detection.

\subsection{Slice Construction}
\label{subsec:query_generation}

\begin{figure}[!ht]
    \centering
    \includegraphics[width=3.42in]{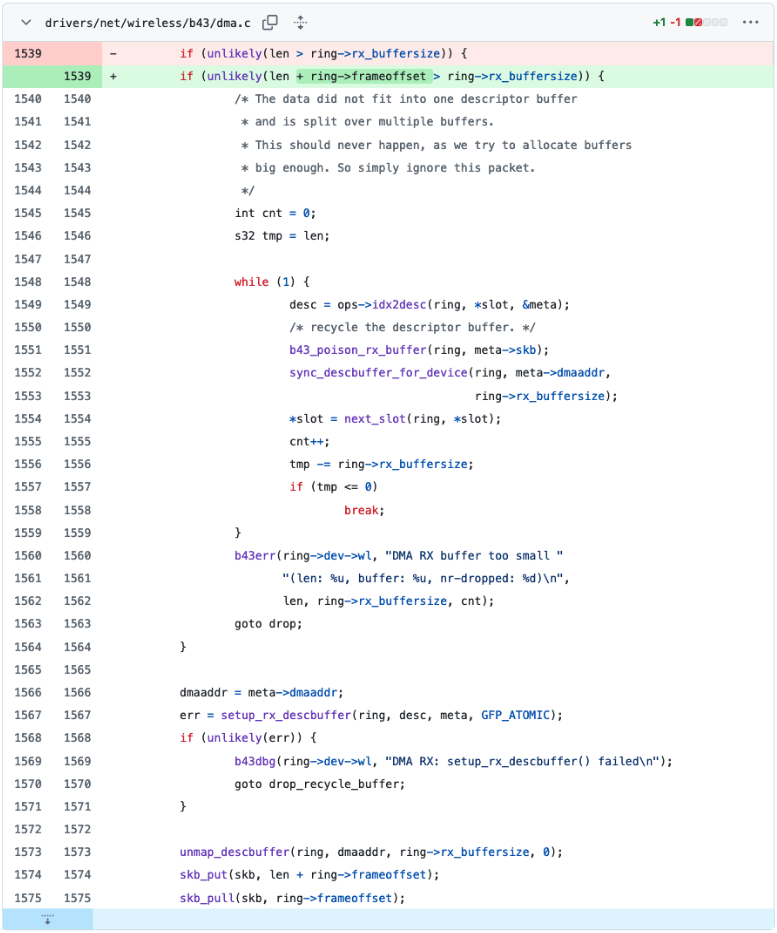}
    \caption{An example of a buffer overflow vulnerability from CVE-2011-3359.}
    \label{fig:example-buffer-overflow}
\end{figure}

\begin{figure}[!ht]
    \centering
    \includegraphics[width=3.42in]{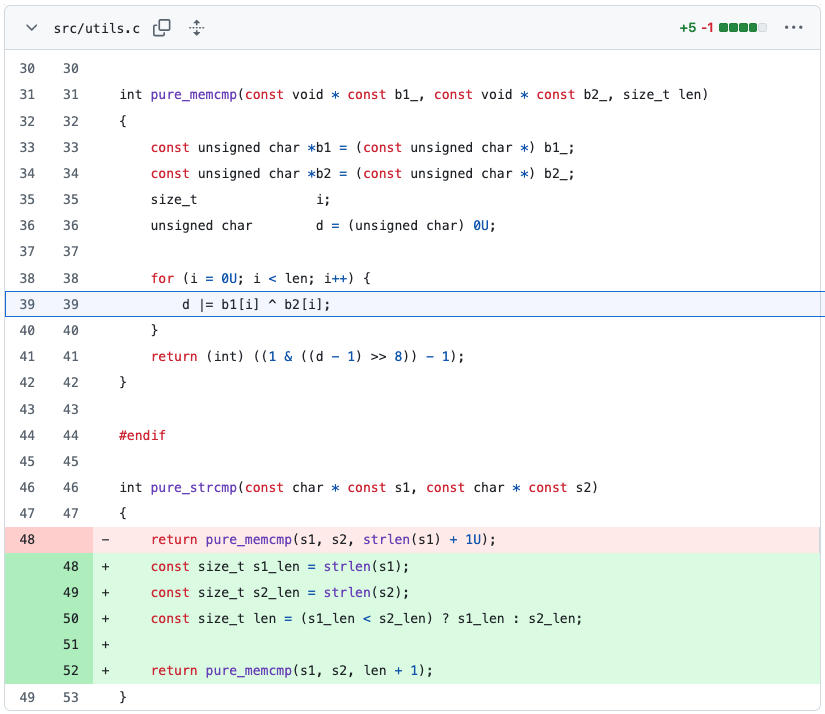}
    \caption{An example of a buffer overflow vulnerability from CVE-2020-9365.}
    \label{fig:example-buffer-overflow-2}
\end{figure}

Current vulnerability detection methods typically analyze code in its raw, unprocessed form, rather than first distilling it down to smaller, more focused code snippets \cite{shimmi2024vulsim}. This approach is problematic because vulnerable code often contains only a small fraction of lines that are actually related to the vulnerability. As a result, detection models face two key challenges: i) including codes irrelevant to vulnerabilities increases token usage, ii) struggling to discern truly relevant vulnerability patterns, often leading to models relying on spurious features \cite{bohme_limits}. 

An intuitive approach to deal with the spurious feature issue is to use program slicing \cite{programslicing}. Program slicing is a method to reduce a program into a smaller representation using data flow and control flow analysis. The outcome of this operation is called a slice, which is an independent program that faithfully represents the original program within the domain of the specified subset of behavior. More specifically, in our case, slices can be utilized  
to capture vulnerability behavior, minimize noise, and focus on the relevant vulnerability features.

To generate a slice, a criterion point must be selected first. In vulnerability detection, the criterion point is a line of code that contains an insecure variable or an insecure function call. A variable or function call is considered insecure when it can potentially lead to security vulnerabilities if not properly handled. This includes user-controlled input that flows into security-sensitive operations without proper validation or sanitization, functions known to be dangerous if misused (like \texttt{strcpy} in C which can cause buffer overflows), or variables that store sensitive data (like passwords or encryption keys) without appropriate protection. These insecure elements are typically identified through a combination of pattern matching against known vulnerable function signatures, taint analysis to track the flow of untrusted data, and control flow analysis to understand how variables and function parameters are used throughout the program. Once a criterion point is identified, the program slice is constructed by analyzing both its dependencies and its impacts: backward slicing captures all code elements that may influence the criterion point's behavior, while forward slicing identifies all code elements that the criterion point may affect. For both operations, the captured code elements include:

\begin{itemize}
    \item \textbf{Data dependencies:} All variables, expressions, and statements that directly or indirectly affect the value of variables used in the criterion point, captured by tracing back the data flow.
    \item \textbf{Control dependencies:} Structures such as \texttt{if}, \texttt{for}, and \texttt{while} statements that determine whether the criterion point is executed, captured by tracing back the control flow.
\end{itemize}

\textbf{Challenges in Program Slicing.} Program slicing is often not straightforward, as it presents various challenges depending on the specific vulnerability being addressed. First, selecting appropriate criterion points is complex: while approaches like UltraVCS \cite{ultravcs} use predefined sets of sensitive function calls in C, developers often create custom wrappers around these functions, making them harder to detect. Other methods, such as Snopy \cite{snopy} and MVP \cite{mvp}, attempt to identify criterion points by analyzing differences between vulnerable and patched code versions, but this approach can be unreliable as patches frequently include unrelated refactoring changes. Second, even when criterion points are correctly identified, the resulting slices often contain unnecessary code. For example, in the buffer overflow vulnerability shown in Figure~\ref{fig:example-buffer-overflow} (CVE-2011-3359), selecting the \texttt{if} condition at line 1539 as the criterion point would include the entire \texttt{if} block in the slice, despite many of these lines being irrelevant to the vulnerability.

To address these challenges, we propose a novel approach that shifts focus from individual criterion points to execution paths. Instead of relying on predefined sensitive functions or code differences, we leverage CPGs to identify potentially vulnerable execution paths. This graph-based approach allows us to capture the essential flow of data and control while minimizing the inclusion of irrelevant code.

\textbf{Slice Construction.} Our slice construction approach works in three main steps. First, we use the SAST tool Joern and its CPGQL query language to identify potential vulnerability root causes in the code, focusing on execution paths rather than analyzing the entire codebases that span multiple functions or files. Second, we analyze each execution path by traversing the CPG to identify variables that interact with that path. Third, we build our final slice by gathering all code elements that influence both the execution path and its interacting variables. The result is a focused code snippet that contains only the essential components: the execution path itself, the variables that interact with it, and any code that affects either of these elements.

\begin{figure}[!ht]
    \centering
    \includegraphics[width=3.3in]{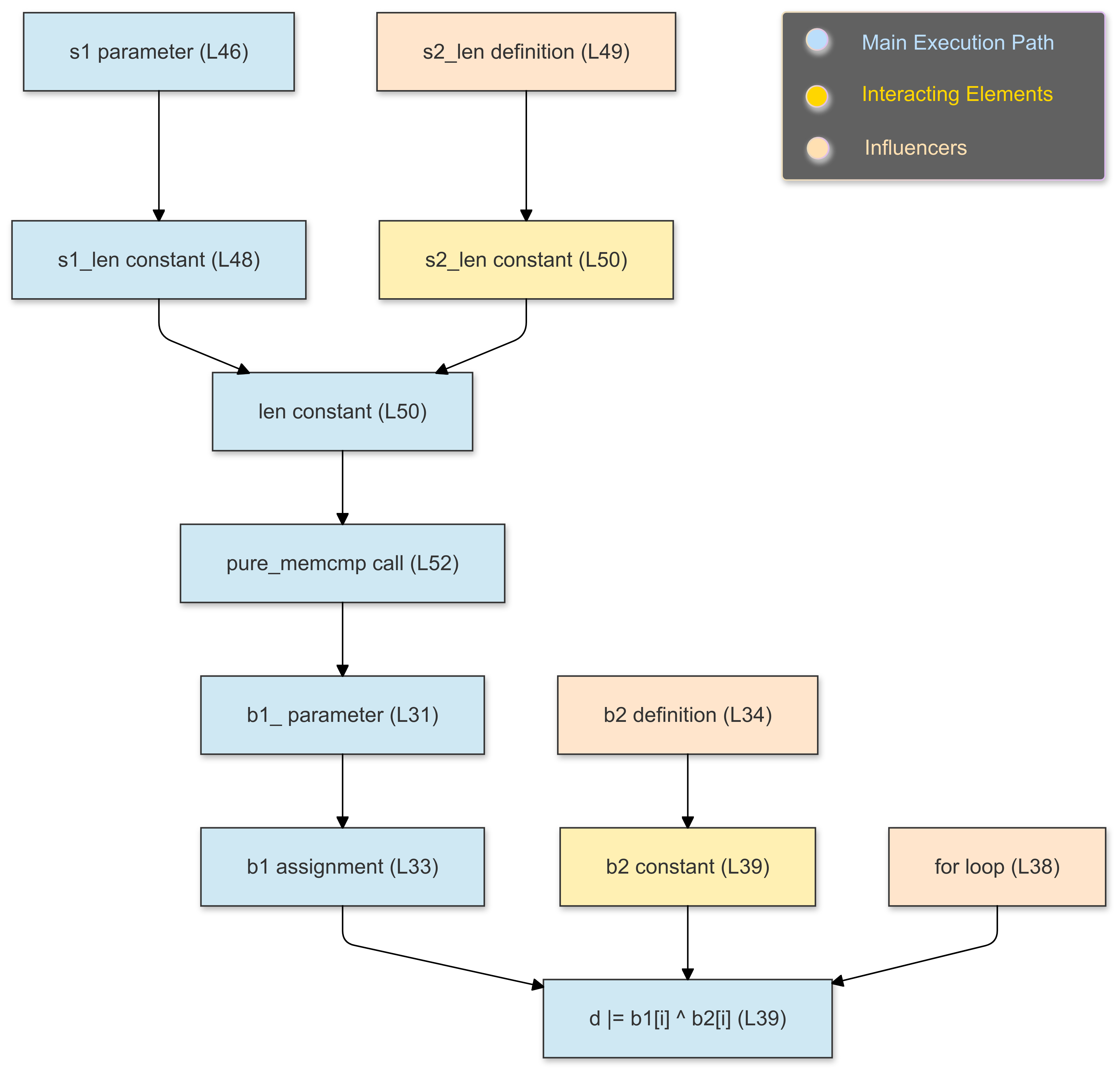}
    \caption{Slice extraction on CVE-2011-3359.}
    \label{fig:slice_extraction}
\end{figure}

\subsubsection{Taint Path Extraction}
\label{subsubsec:taint_path_extraction}

\begin{lstlisting}[language=Scala, caption={Example CPGQL queries to identify sources, sinks, and execution paths.}, label={lst:sourceSinkPaths}, float=t]
val source = cpg.identifier.name("len")
val sink = cpg.call.name("skb_put").where(_.argument.order(2).codeExact("len + ring->frameoffset"))
val execution_paths = sink.reachableByFlows(source)
\end{lstlisting}

For taint path extraction, we employ CPGQL, a specialized query language designed for analyzing code property graphs in Joern. CPGQL facilitates the navigation and analysis of code property graphs via queries targeting specific code patterns. For instance, the following query identifies all function calls within a method named \texttt{processData}:
\begin{lstlisting}[language=Scala]
cpg.method.name("processData").call.name
\end{lstlisting}

A critical step in \system is the design of CPGQL queries to extract execution paths relevant to the target vulnerabilities. For example, in Listing \ref{lst:sourceSinkPaths}, the first query identifies the source of the vulnerability (the identifier \texttt{len}), while the second locates the sink (a call to \texttt{skb\_put}), which might fail without proper buffer size checks. The final query identifies all execution paths between the source and sink.

\textbf{Fine Tuning for Query Generation.} The goal of this step is to \textcolor{black}{to have a model that can generate valid CPGQL queries that target a specific vulnerability pattern depending only on the provided code snippet with no additional information such as CWE type, or vulnerability location}.
While powerful language models like DeepSeek, ChatGPT, and Qwen excel at general code generation, they initially struggle with generating effective CPGQL queries since CPGQL is a low-resource language. However, since CPGQL is based on Scala, we leverage this similarity to fine-tune the Qwen2.5-Coder-32B-Instruct model (currently the best code model available) to generate CPGQL queries. To create our training data, we utilize DeepSeek-v3 to generate an initial set of queries. We generate and test queries on a Joern server, Feedback is provided to the model whenever queries contain syntax errors or fail to identify vulnerable paths, enabling iterative refinement and improvement. Henceforth, we will refer to the query generation fine-tuned model as \systemquery.

\subsubsection{Interacters: Finding Variables that Interact with the Execution Path}
\label{subsubsec:interacters}

As shown in Figure~\ref{fig:slice_extraction}, each extracted path represents a specific flow of data and control through the code. For example, in the CVE-2020-9365 vulnerability (Figure \ref{fig:example-buffer-overflow-2}), \system identifies a key execution path (shown in Blue Boxes).
However, this execution path alone does not provide a complete vulnerability assessment, as it lacks the definition of the \texttt{s2} constant. Without this contextual information, determining the security implications of this execution path is not feasible. 
To build a comprehensive understanding, we must identify all code elements that interact with this path. We achieve this by navigating the CPG using the query language to find all interacting code elements. Listing \ref{lst:interacters_query} shows two queries. The first query, generated by \systemquery, captures potential vulnerable execution paths. The second query identifies all identifiers that interact with the captured execution path. An identifier is considered an interacter if its line number matches the line number of at least one code element in the execution path.

\begin{lstlisting}[language=Scala, caption={CPGQL query to identify nodes that interact with the extracted execution path.}, label={lst:interacters_query}]
val execution_path_nodes = <query to extract the execution path generated by LLMxCPG-Q>
cpg.identifier.filter(id => execution_path_nodes.lineNumber.toSet.intersect(id.lineNumber.l.toSet).size.equals(1))
\end{lstlisting}

\subsubsection{Backward Slicing for Focused Code Snippet Construction}
\label{subsubsec:backward_slice}

In the final step, we use backward slicing to build a complete code snippet that includes both the execution path and its interacting elements. As illustrated in Figure \ref{fig:slice_extraction}, this process captures all essential dependencies, such as the \texttt{s2$\textunderscore$len} definition, the \texttt{b2} string initialization, and the relationship between the \texttt{for loop} and the \texttt{len} variable. This comprehensive slice provides all the context needed to understand the potential vulnerability. We note that we automate the process of applying a backward slice using a CPGQL query. Listing \ref{lst:backward_slice_query} shows the query used to perform backward slicing. The \texttt{reachableByFlows} API identifies all code elements that influence either the execution path or its interacters. Internally, Joern utilizes the Program Dependency Graph (PDG) to construct the backward slice.

\begin{lstlisting}[language=Scala, caption={CPGQL query to apply backward slicing.}, label={lst:backward_slice_query}]
... queries to extract execution path and the interacters.
execution_path_and_interacters.reachableByFlows(cpg.all)
\end{lstlisting}

\begin{figure*}
    \centering
    \includegraphics[scale=0.13]{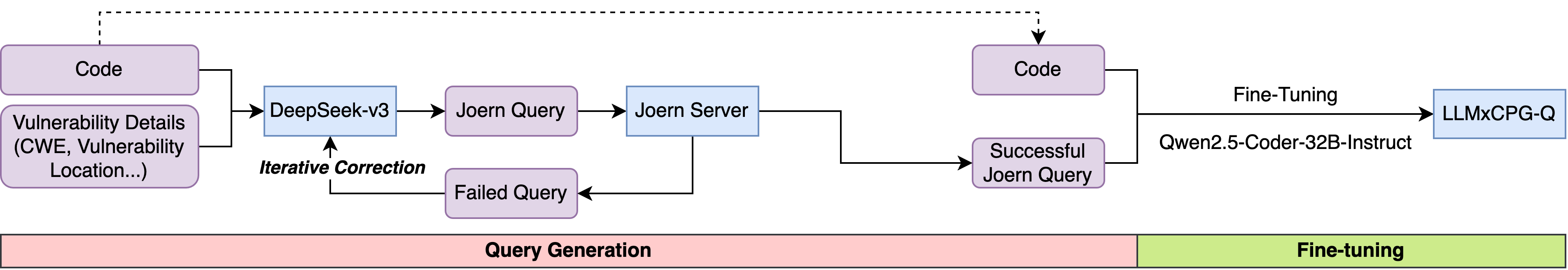}
    \caption{Query Generation Workflow}
    \label{fig:query_generation_workflow}
\end{figure*}

\subsection{Vulnerability Detection}
\label{subsec:vulnerability_detection}

\begin{figure*}
    \centering
    \includegraphics[scale=0.13]{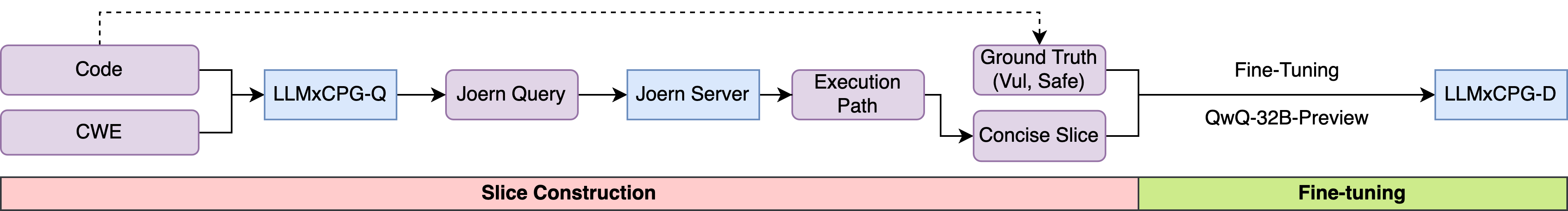}
    \caption{Code Classification Workflow}
    \label{fig:vul_detection_workflow}
\end{figure*}

The slice construction process reduces code samples that span multiple functions and multiple files to concise snippets of code that better represent the characteristics of the vulnerabilities. For example, the 85 lines of code function in Figure \ref{fig:example-buffer-overflow}, can be compressed to an 18-line code function by using CPGQL queries as shown in Listing \ref{lst:path-code-snippet}. This path extraction approach significantly enhances vulnerability detection by isolating the security-relevant code patterns specific to each CWE type while eliminating non-essential code. By focusing on the critical data and control flow paths that characterize potential vulnerabilities (such as source-to-sink paths for taint-style vulnerabilities or validation-check patterns for input handling flaws), this method minimizes noise that would otherwise obscure vulnerability signatures. We show the overall workflow to classify code as \textit{Vulnerable} or \textit{Safe} in Figure \ref{fig:vul_detection_workflow}.

\textbf{Fine Tuning for Classification.} As previously noted, we fine-tuned the QwQ-32B-Preview model for the classification task. To construct the fine-tuning dataset for classifying code slices as either vulnerable or safe, we extracted code snippets from both vulnerable and safe samples in the training dataset using \systemquery given the groundtruth labels in our training datasets.
To this end, the fine-tuned classifier model will be referred to as \systemclassifier.

\begin{lstlisting}[language=C, caption={CPGQL captured path represented as a code snippet.}, label={lst:path-code-snippet}, float=t]
static void dma_rx(struct b43_dmaring *ring, int *slot)
{
    u16 len;
    len = le16_to_cpu(rxhdr->frame_len);
    if (unlikely(len > ring->rx_buffersize)) {
        s32 tmp = len;
        while (1) {
            tmp -= ring->rx_buffersize;
            if (tmp <= 0)
                break;
        }
        goto drop;
    }

    skb_put(skb, len + ring->frameoffset);
    drop:
    return;
}
\end{lstlisting}

Note that this approach is equally effective for compressing safe code samples. For instance, in the previously mentioned example in Figure \ref{fig:example-buffer-overflow}, the patched version of the code involves only a single change at line 1539, where the if condition was modified to ensure sufficient buffer size. Applying the same CPGQL query-based approach to this patched version reduces noise and highlights the specific security-relevant modification. The ability to precisely identify and isolate security-critical changes between vulnerable and patched versions makes \system particularly valuable for understanding vulnerability fixes and generating high-quality training data for vulnerability detection models. In fact, since high-quality datasets in the vulnerability detection domain are scarce, we plan to use our approach in future work to compile an open-source, large, and high-quality dataset for training vulnerability detection models. 

\section{Evaluation}
\label{sec:evaluation}

This section presents the employed datasets and shows details of our implementation. In addition, we provide a detailed analysis of \system performance on both function-level and project-level datasets. Moreover, we analyze its ability to generalize to unseen datasets and its robustness against code transformations as defined in \cite{bohme_limits}.

\subsection{Datasets}
\label{subsec:datasets}

\textbf{Training Datasets.} We used two datasets for training, FormAI-v2 \cite{formai-v2} and PrimeVul \cite{primevul}. The FormAI-v2 dataset includes 331,000 compilable C programs generated using various LLMs including Google's GEMINI-pro, OpenAI's GPT-4, TII's 180 billion-parameter Falcon, CodeLLama2, and other compact models. These programs are generated using a dynamic zero-shot prompting technique and comprise programs with varying levels of complexity. Each program is labeled for code vulnerabilities using a formal verification method based on the Efficient SMT-based Bounded Model Checker (ESBMC) \cite{esbmc}. FormAI minimizes false negatives by ensuring comprehensive formal verification of the code within a defined timeframe.
We note that FormAI C programs 
are not directly mapped to CWEs, but rather correspond to errors returned by ESMBC. To address this, we manually created a mapping of error messages (provided by ESBMC) to CWEs.
On the other hand, the PrimeVul dataset was created to address the shortcomings of existing vulnerability datasets, such as poor data quality, low label accuracy, and high duplication rates. PrimeVul employs novel data labeling techniques, achieving a label accuracy comparable to human-verified benchmarks, while significantly expanding the dataset. The dataset implements rigorous data de-duplication and chronological data splitting to avoid data leakage issues. The dataset contains 228,800 safe functions and 6,968 vulnerable functions covering 140 CWEs, making it a diverse and accurate resource for vulnerability detection research. The dataset is split chronologically based on commit dates, with 80$\%$ for training, 10$\%$ for validation, and 10$\%$ for testing.

\textbf{Generalizability.} To evaluate generalizability, we utilize two distinct vulnerability datasets: SVEN \cite{sven} and ReposVul \cite{wang2024reposvul}. SVEN is a manually curated dataset of approximately 1,600 C/C++ and Python programs derived from real-world GitHub security fixes, with a rigorous verification process to ensure high data quality and relevance to security hardening. ReposVul complements this with a broader repository-level perspective, encompassing 6,134 CVE entries across 236 CWE types from 1,491 projects in four programming languages. Together, these datasets provide a comprehensive foundation for assessing vulnerability detection capabilities across different real-world scenarios.

\textbf{Studied CWEs.} Our analysis focuses on a specific subset of memory-related CWEs that are amenable to static analysis via Code Property Graphs (CPGs): CWE-119 (Buffer Overflow), CWE-190 (Integer Overflow), CWE-415 (Double Free), and CWE-416 (Use After Free). Additionally, we include CWE-120 (Buffer Copy without Checking the Size of Input), CWE-121 (Stack-based Buffer Overflow), CWE-122 (Heap-based Buffer Overflow), CWE-125 (Out-of-bounds Read), and CWE-787 (Out-of-bounds Write), which are specialized variants of CWE-119. This selection criterion balances the prevalence of real-world vulnerabilities with the technical constraints of CPG-based detection. We specifically excluded vulnerabilities that rely on dynamic program behavior, such as race conditions, as these cannot be reliably modeled through static CPG analysis due to their runtime-dependent nature.

\textbf{CWE Distribution.} \khang{The CWE distribution varies across our datasets, as illustrated in Table \ref{tab:cwe_distribution}.} FormAI-v2's training set contains 5,893 vulnerable samples across four CWE types (CWE-119, CWE-190, CWE-415, and CWE-416), with relatively balanced distribution ranging from 1,395 to 1,500 samples per CWE, alongside 4,431 safe samples. \khang{It is worth noting that the FormAI-v2 dataset was validated using ESBMC with default memory violation assertions and does not provide paired vulnerable–patched code samples, resulting in non-indication of safe samples per CWE.} 
\textcolor{black}{In other words, the safe samples were formally verified to be free from memory violations under the default ESBMC configuration. To this end, we exclude samples with unknown or timed-out verification results, as these do not conclusively demonstrate the presence or absence of a vulnerability. Regarding CWE mapping for the safe samples, this does not apply to FormAI-v2 because, unlike real-world cases where vulnerable and patched versions of the same code are available, the code snippets in FormAI-v2 are generated independently.}
PrimeVul's training data spans eight CWE types, with CWE-119 being the most prevalent (518 samples) and some CWEs having limited representation (e.g., CWE-121 and CWE-122 with only 1-2 samples). For our test datasets, we ensure balanced representation where possible. SVEN's test set contains four CWE types (CWE-125, CWE-190, CWE-416, and CWE-787) with paired vulnerable and safe samples ranging from 37 to 122 pairs per CWE. Similarly, we select a balanced subset from ReposVul, including matched safe and vulnerable samples across seven CWE types.

\begin{table}[!htpb]
    \centering
    \caption{CWE Distribution Across Datasets}
    \vspace{2pt}
    \label{tab:cwe_distribution}
    \resizebox{\columnwidth}{!}{%
    \begin{threeparttable}
    \begin{tabular}{lcccccccc}
    \toprule
    \multirow{2}{*}{\textbf{CWE}} & \multicolumn{3}{c}{\textbf{Training Datasets}} & \multicolumn{5}{c}{\textbf{Test Datasets}} \\
    \cmidrule(rl){2-4} \cmidrule(rl){5-9}
     & FormAI & PrimeVul & Total & SVEN & ReposVul & FormAI & PrimeVul & Total \\
    \midrule
    CWE-119 & 1,395\textcolor{black}{/NA} & 518\textcolor{black}{/518} & 1,913\textcolor{black}{/518} & -- & 19\textcolor{black}{/19} & 51\textcolor{black}{/NA} & -- & \textcolor{black}{70/19} \\
    CWE-120 & -- & 35\textcolor{black}{/35} & 35\textcolor{black}{/35} & -- & 4\textcolor{black}{/3} & -- & --\textcolor{black}{/2} & \textcolor{black}{4/5} \\
    CWE-121 & -- & 1\textcolor{black}{/1} & 1\textcolor{black}{/1} & -- & 1\textcolor{black}{/1} &  -- & -- & \textcolor{black}{1/1} \\
    CWE-122 & -- & 2\textcolor{black}{/2} & 2\textcolor{black}{/2} & -- & 2\textcolor{black}{/--} &  -- & -- & \textcolor{black}{2/--} \\
    CWE-125 & -- & 391\textcolor{black}{/391} & 391\textcolor{black}{/391} & 122\textcolor{black}{/122} & 13\textcolor{black}{/19} & -- & 9\textcolor{black}{/6} & \textcolor{black}{144/147} \\
    CWE-190 & 1,500\textcolor{black}{/NA} & 138\textcolor{black}{/138} & 1,638\textcolor{black}{/138} & 37\textcolor{black}{/37} & 4\textcolor{black}{/3} & 41\textcolor{black}{/NA} & 11\textcolor{black}{/12} & \textcolor{black}{93/52} \\
    CWE-415 & 1,499\textcolor{black}{/NA} & 49\textcolor{black}{/49} & 1,548\textcolor{black}{/49} & -- & 4\textcolor{black}{/3} & 50\textcolor{black}{/NA} & 8\textcolor{black}{/15} & \textcolor{black}{62/18} \\
    CWE-416 & 1,499\textcolor{black}{/NA} & 176\textcolor{black}{/176} & 1,675\textcolor{black}{/176} & 56\textcolor{black}{/56} & 13\textcolor{black}{/12} & 43\textcolor{black}{/NA} & 12\textcolor{black}{/5} & \textcolor{black}{124/73} \\
    CWE-787 & -- & -- & -- & 44\textcolor{black}{/44} & -- & -- & -- & \textcolor{black}{44/44} \\
    \midrule
    Total Vulnerable & 5,893 & 1,310 & 7,203 & 259 & 60 & 185 & 40 & 544 \\
    Total Safe & 4,431 & 1,310 & 5,741 & 259 & 60 & 198 &  40 & 557 \\
    \bottomrule
    \end{tabular}
    \begin{tablenotes}
      \small
      \item \textcolor{black}{Note: For the PrimeVul, SVEN, and ReposVul datasets, numbers are shown as vulnerable/safe pairs. '--' indicates absence of the CWE type. For FormAI dataset, the samples are not paired, thus only the total safe samples are stated.}
    \end{tablenotes}
    \end{threeparttable}
    }
\end{table}

\subsection{Implementation}
\label{subsec:implementation}

Our implementation has three main components: slice construction, fine-tuning, and inference. 

\textbf{Slice Construction.} In this step, we train the CPGQL query generation model using a training dataset of valid Joern queries. To construct these training queries, we leverage DeepSeek v3 to generate potential candidates, which are then validated using Joern, as illustrated in Figure \ref{fig:query_generation_workflow}.
Failed queries are returned to DeepSeek along with the error message generated by Joern for up to two additional attempts. If the query remains invalid after three attempts, it is discarded. 
We deploy a cluster of Joern servers using Docker containers to enable parallel processing of multiple repositories. 



\textbf{Finetuning.} After extensive evaluation of state-of-the-art language models including Phi-4, Qwen2.5-Coder-32B-Instruct, and Codestral 22B, we selected QwQ-32B-Preview as our base model for fine-tuning \systemclassifier and Qwen2.5-Coder-32B-Instruct for finetuning \systemquery \textcolor{black}{(see Appendix \ref{apdx:comparison_with_base_models} for more details)}. For both models, the fine-tuning process employs Low-Rank Adaptation (LoRA) \cite{hu2021lora} using the LLaMA-Factory\footnote{\url{https://github.com/hiyouga/LLaMA-Factory}} framework. We configure the LoRA parameters with rank 8 and alpha 4, while setting the learning rate to $10^{-4}$. The fine-tuning is performed on an NVIDIA A100-80GB GPU running Ubuntu 22.04, allowing us to effectively adapt the pre-trained model to the specific task of vulnerability detection.

\textbf{Inference Pipeline.} For query inference, we employ the inference library vLLM\footnote{\url{https://github.com/vllm-project/vllm}} with a CPGQL generation prompt shown in Appendix \ref{apdx:promt_templates} (Figure \ref{fig:query_gen_prompt}). For classification inference, we employ Unsloth\footnote{\url{https://docs.unsloth.ai}}, a library for efficient finetuning and inference. We implement an optimized pipeline that reduces the \systemclassifier's  language modeling head to focus exclusively on binary classification between vulnerable and safe code samples. Specifically, we extract the weight vectors corresponding to the \textit{"Yes"} (i.e., Vulnerable) and \textit{"No"} (i.e., Safe) tokens from the original \texttt{lm\_head}, constructing a reduced classification head. During inference, \systemclassifier generates logits for these two tokens, which are then passed through a softmax function to obtain prediction probabilities. This approach significantly reduces the computational overhead compared to full token generation while maintaining classification accuracy. A threshold $\gamma$ is applied to these probabilities to determine the final classification. 

The classification pipeline can be formally described as follows: Let $w_v$ and $w_b$ represent the weight vectors corresponding to \textit{Vulnerable} and \textit{Safe} tokens respectively, extracted from the original language model head $W_{\text{lm}}$. The classification logits $\mathbf{l}$ for an input code sample $x$ are computed as:

\[
\mathbf{l} = [l_v, l_b] = f_\theta(x) \cdot [w_v, w_b]^T
\]

 where $l_v$ is the logit corresponding to \textit{Vulnerable}, $l_b$ is the logit corresponding to \textit{Safe}, and $f_\theta(x)$ represents the model's output. These logits are transformed into probabilities through the softmax function:

\[
p(y \mid x) = \text{softmax}(\mathbf{l}) = \left[\frac{e^{l_v}}{e^{l_v} + e^{l_b}}, \frac{e^{l_b}}{e^{l_v} + e^{l_b}}\right]
\]

The final classification is determined by comparing the vulnerability probability against a dataset-specific threshold $\gamma$:

\[
\hat{y} =
\begin{cases} 
1 & \text{if } p(y = \text{vulnerable} \mid x) > \gamma \\
0 & \text{otherwise}
\end{cases}
\]

\khang{The threshold $\gamma$ can be predefined for the system as a hyper-parameter. Furthermore, it can be selected by security analysts based on their small validation set of known vulnerabilities, which is practical and facilitates system adaptation without extensive manual labeling.}
Our empirical analysis, shown in Figure \ref{fig:threshold_accuracy_analysis} revealed that threshold calibration is essential for optimal performance across different datasets. We determine distinct threshold values for each dataset in our study through validation splits. Notably, we observe higher threshold values for datasets used in training unlike unseen datasets.
This pattern suggests the model expresses higher confidence when evaluating code patterns similar to its training distribution. For users applying our model to new datasets, we recommend calibrating the threshold using a small random sample (as few as 20 data points) of labeled examples from their target domain. We use the following thresholds after sampling 20 data points from validation splits of the employed datasets:  PrimeVul $\gamma = 0.594$, FormAI: $\gamma = 0.547$, SVEN: $\gamma = 0.334$, and ReposVul: $\gamma = 0.193$. We show the effect of vaying thresholds on \systemclassifier's accuracy in Appendix \ref{apdx:choose_threshold}.


\subsection{Performance Analysis}
\label{subsec:eval_performance_analysis}

\subsubsection{Query Generation}
\label{subsubsec:eval_query_generation}


\begin{table}[t]
    \centering
    \caption{Number of valid queries per model out of 1278 test samples.}
    \label{tab:number_of_valid_queries}
    \resizebox{\columnwidth}{!}{%
    \begin{threeparttable}
    
    \begin{tabular}{lc}
    \toprule
    \textbf{Model} & Number of valid queries \\
    \midrule
    DeepSeek-v3 & 132 \\ 
    Qwen2.5-Coder-32B-Instruct & 19 \\ 
    \systemquery & 1278 \\ 
    \bottomrule
    \end{tabular}
    \end{threeparttable}
    }
\end{table}


As shown in Table \ref{tab:number_of_valid_queries}, \systemquery effectively learns the syntax of CPGQL, whereas the base models, Qwen2.5-Coder-32B-Instruct and DeepSeek-v3, encounter difficulties in generating valid CPGQL queries. To further investigate the quality differences among the studied models, we randomly selected 50 samples where DeepSeek generated invalid queries to analyze what \systemquery had successfully learned. We classify the invalid queries into the following categories:
\begin{itemize}
    \item \textbf{Incorrect usage of CPGQL APIs:} DeepSeek often misuses the \texttt{.code} API to filter nodes by their name. For instance, to retrieve all \texttt{Call} nodes with the name "\texttt{print}", DeepSeek generates the query \texttt{cpg.call.code("print")}. However, this query returns empty results because the \texttt{code} property matches the entire statement, including the arguments of the function. The correct query for this scenario is \texttt{cpg.call.name("print")}, which matches only the function's name.
    \item \textbf{Missing or incorrect API usage:} DeepSeek mixes node types with operations allowed on each type. For example, it generated the query \texttt{val inputSources = cpg.call.code("scanf").argument.filter(\_ .typeFullName.matches("float")).toList}, where \texttt{typeFullName} is incorrectly applied to \texttt{argument} nodes.
    \item \textbf{Incorrect handling of regex-supporting APIs:} Some CPGQL filtering APIs support regular expressions, but DeepSeek fails to distinguish between these APIs. For example, it may use \texttt{cpg.call.code('a + b')} to search for an addition operation, which fails due to regex interpretation. The correct query for this case is \texttt{cpg.call.codeExact('a + b')}, which performs an exact string match.
\end{itemize}

\revision{
While our evaluation demonstrates that \systemquery successfully learns the intricacies of CPGQL syntax—including proper API usage and regex handling—syntactic correctness alone is insufficient for real-world applications. A truly effective system must also maintain semantic precision, ensuring the generated queries accurately isolate vulnerability patterns within the code.

To assess this semantic dimension, we conducted a rigorous human-auditing process involving three security experts who manually evaluated 50 queries generated by \systemquery. These queries were sampled from the PrimeVul and SVEN test datasets and included 25 true positive/negative and 25 false positive/negative cases, \textcolor{black}{ where the ground truth labels (true/false positive/negative) were determined by the final classification decisions of \systemclassifier on the constructed slices (e.g., our evaluation considers a case as positive if \systemclassifier correctly labels the constructed slice, even when the generated query targets a different CWE than the actual vulnerability's CWE type)}. Each expert assessed whether the queries \textcolor{black}{are semantically} aligned with the intended vulnerability patterns and successfully captured execution paths that isolate the vulnerabilities with minimal noise. In 76\% of the true positive/negative samples, the queries matched the vulnerability pattern semantically and yielded meaningful paths. The experts achieved a Fleiss’ Kappa \cite{fleiss1971measuring} score of 0.6429 \textcolor{black}{across the 25 true positive/negative samples}, indicating substantial inter-rater agreement according to the guidelines provided by Landis and Koch \cite{landis1977measurement}. Discrepancies were resolved through discussion.
To better understand the limitations, we analyzed the 25 false positive/negative samples and categorized them as follows:
\begin{itemize}
    \item 28\% of the queries were semantically correct and accurately identified the vulnerability pattern, despite being labeled incorrectly.
    \item 40\% of the queries targeted a different CWE, failing to capture the correct vulnerability.
    \item 32\% correctly identified the CWE but missed critical contextual elements necessary for a complete vulnerability match.
\end{itemize}

Although \systemclassifier misclassifies certain cases even when a vulnerability is present in the constructed code slice (28\% of false positives/negatives and 24\% of true positives/negatives), most of these errors occur when the query model fails to detect the target vulnerability pattern.
}

\textcolor{black}{
In Appendix \ref{apdx:impact_of_slice_construction_and_query_generation_model}, we evaluate the performance of \systemclassifier on full code snippets (omitting the slicing step). We then run Joern-scan\footnote{\url{https://docs.joern.io/scan/}}, a tool that executes predefined queries targeting various vulnerability patterns, on a subset of the test set to gauge the impact of the slice construction and query generation steps.
}

\textbf{Code Reduction.} Our empirical analysis reveals significant code reduction ratios achieved through our slice construction approach across multiple datasets. In the synthetic FormAI dataset, we observe an average reduction of 78.70\% in code size on the test set. Similar efficiency is demonstrated in real-world scenarios, with code reductions of 67.84\% and 70.22\% in the function-level datasets PrimeVul and SVEN respectively, and a substantial 90.93\% reduction in the project-level dataset ReposVul. 
In the subsequent section, we conduct a comprehensive analysis of \system's performance on these reduced code slices to examine whether the system effectively capitalizes on these significant reductions while maintaining vulnerability detection capabilities.


\subsubsection{Function-level Vulnerability Detection}
\label{subsubsec:eval_vuln_detection}

Using the FormAI and PrimeVul datasets, we evaluate \system's performance in detecting vulnerabilities across diverse function-level code snippets. These datasets provide a comprehensive testbed, capturing various vulnerability types and coding practices. 

\begin{table}[!htbp]
    \centering
    \caption{Average Performance of \system on PrimeVul \& FormAI datasets}
    \label{tab:performance-on-formAI-primevul}
    
    \resizebox{\columnwidth}{!}{
    \begin{tabular}{lcccc}
    \toprule
     Dataset & Accuracy & Precision & Recall & F1-score \\
     \midrule
     FormAI & 0.8146 & 0.8097 & 0.8054 & 0.8075 \\ 
     PrimeVul & 0.7250 & 1.0 & 0.45 & 0.6206 \\ 
    \bottomrule
    \end{tabular}
    }
\end{table}

Table \ref{tab:performance-on-formAI-primevul} illustrates the results of \system. In general, \system achieves high performance on vulnerability detection in function-level code snippets. Specifically, \system reaches up to 0.8146 Accuracy and 0.8075 F1-score on the FormAI dataset. Similar results are observed on the PrimeVul dataset, which indicates the effectiveness of \system in vulnerability detection at the function~level.


\begin{table}[h]
\caption{A Breakdown of the Performance of \system by CWEs on PrimeVul \& FormAI datasets}
\label{tab:cwe_metrics}
\centering
\resizebox{\columnwidth}{!}{
\tiny
\begin{tabular}{lcccc}
\toprule
\textbf{CWE} & \textbf{Accuracy} & \textbf{Precision} & \textbf{Recall} & \textbf{F1-Score} \\
\midrule
CWE-119 & 0.941 & 1.000 & 0.941 & 0.970 \\
CWE-415 & 0.757 & 0.891 & 0.817 & 0.852 \\
CWE-416 & 0.778 & 1.000 & 0.736 & 0.848 \\
CWE-190 & 0.672 & 0.844 & 0.745 & 0.792 \\
\bottomrule
\end{tabular}
}
\end{table}

Table \ref{tab:cwe_metrics} illustrates the performance of \system with respect to different CWEs. \system~exhibited strong performance on several memory-related CWEs. Specifically, for CWE-119, the model achieved significantly high accuracy and F1-scores (0.941 and 0.97, respectively). Similar results are also observed in other CWEs. 



\subsection{Generalizability}
\label{subsec:eval_generalizability}

\subsubsection{Function-level Vulnerability Detection}
\label{subsubsec:eval_func_level_gen}


This section assesses the generalizability of \system for detecting vulnerabilities in function-level code snippets. To provide a comprehensive evaluation, we benchmark its performance on real-world code snippets against a range of state-of-the-art baselines, including VulSim \cite{shimmi2024vulsim}, ReGVD \cite{nguyen2022regvd}, and both VulBERTA-CNN and VulBERTA-MLP \cite{hanif2022vulberta}. VulSim combines the structural and semantic information similarity of a function-level code snippet with the snippets in the training datasets to classify whether it is vulnerable or safe. Similarly, VulBERTA models incorporate the CodeBERT transformers model to extract the semantics of the snippets to make classifications. In addition, ReGVD leverages the graph structural information of the code snippets to classify whether it is vulnerable or not. To ensure a fair comparison, we employ SVEN dataset as it was not part of the training data of \system and the considered baselines.


\begin{table}[!htbp]
    \centering
    \caption{Function-level vulnerability detection average performance on SVEN dataset, which includes CWE-125, CWE-190, CWE-416, CWE-476.}
    \label{tab:performance-llmxcpg}
    \resizebox{\columnwidth}{!}{%
    \begin{threeparttable}
    
    \begin{tabular}{lcccc}
    \toprule
     & Accuracy & Precision & Recall & F1-score \\
    \midrule
    VulSim \cite{shimmi2024vulsim} & 0.33 & 0.31 & 0.31 & 0.31 \\ 
    VulBERTA-CNN \cite{hanif2022vulberta} & 0.5 & 0.51 & 0.38 & 0.44 \\ 
    VulBERTA-MLP \cite{hanif2022vulberta} & 0.5 & 0.5 & 0.37 & 0.43 \\ 
    ReGVD \cite{nguyen2022regvd} & 0.51 & 0.53 & 0.46 & 0.55 \\
    \system & \textbf{0.6020} & \textbf{0.5590} & \textbf{0.9534} & \textbf{0.7048} \\
    \bottomrule
    \end{tabular}
    \end{threeparttable}
    }
\end{table}

Table \ref{tab:performance-llmxcpg} presents the vulnerability detection performance of \system on the SVEN dataset, offering insights into its effectiveness on previously unseen code snippets. The results indicate that \system significantly outperforms state-of-the-art baselines in detecting vulnerabilities. In particular, \system achieves a remarkable 20\% improvement in accuracy over the competing approaches. This substantial margin highlights its ability to generalize across different code snippets and quickly uncover vulnerabilities. Such performance underscores the robustness of \system in scenarios where training data does not directly overlap with the test set, further solidifying its value in practical, real-world applications. The superior performance of \system can be attributed to its key design features. One of the critical factors is its integration of CPGs, which allows it to understand the semantics and structure of the code. Additionally, \system leverages specialized LLMs, enabling it to effectively capture vulnerability signatures and adapt to unseen scenarios.


\subsubsection{Project-level Vulnerability Detection}
\label{subsubsec:eval_project_level_gen}

Project-level vulnerability detection presents fundamentally different challenges compared to function-level analysis in terms of code complexity. To comprehensively evaluate the model's performance, we consider five complexity metrics: Lines of Code (LOC), Cyclomatic Complexity (CC), Number of Functions, Number of Branches, and Nesting Depth. These metrics collectively capture different dimensions of code complexity, from pure size (LOC) to structural intricacy (CC, Nesting) and modularity (Functions, Branches). We formally define these metrics in Appendix \ref{apdx:code_metrics}. 

We evaluate \system on a sampled dataset from ReposVul \cite{wang2024reposvul}, despite it not being trained on project-level real-world data. The sampled dataset from ReposVul \cite{wang2024reposvul} comprises 120 balanced samples from 53 real-world projects with complex code snippets spanning multiple files and multiple functions as evidenced by its code metrics statistics shown in Table \ref{tab:reposvul_dataset_stats}.  
On average, each file contains 659 lines of code (LOC), 102.91 branches, 9.28 functions, and a nesting depth of 4.62 levels. These metrics highlight the increased complexity of project-wide vulnerability detection compared to function-level analysis. For instance, while the average number of functions per file is 9.28, a single file may contain up to 59 functions interacting across 335 branch points, substantially exceeding the complexity seen in isolated function analysis. 
Additionally, the high average nesting depth (4.62 levels, with a maximum of 11) underscores the challenge of identifying vulnerabilities that may emerge from deeply nested and interdependent code structures. Such complexity demonstrates the inadequacy of state-of-the-art models limited to function-level detection. These models fail to capture the dependencies and interactions that span multiple functions and deeply nested branches within a project. The ReposVul dataset thus provides a challenging benchmark for advancing project-wide vulnerability detection models capable of addressing these challenges.

\begin{table}[!htpb]
\caption{Code Metrics Statistics of the Sampled ReposVul Dataset}
\label{tab:reposvul_dataset_stats}
\centering
\footnotesize
\begin{tabularx}{\linewidth}{XCCC}
\toprule
Metric & Mean & Min & Max \\
\midrule
LOC       & 659.06 & 54.00  & 1951.00 \\
CC        & 101.71 & 3.00   & 436.00  \\
Functions & 9.28   & 0.00   & 59.00   \\
Branches  & 102.91 & 0.00   & 335.00  \\
Nesting   & 4.62   & 2.00   & 11.00   \\
\bottomrule
\end{tabularx}
\end{table}

\textbf{Performance Analysis on ReposVul dataset.} Despite these challenges, \system~achieves promising results with an average Accuracy of 0.634 and an F1-score of 0.610 on ReposVul. Detailed analysis across complexity metrics reveals interesting patterns. Performance is not affected by the LOC as \system~achieves 0.83 accuracy on samples with extremely high LOC (1623 mean), demonstrating robustness to code size variations. 
It shows strong performance on samples with high cyclomatic complexity (0.75 Accuracy for CC 115-150), indicating effective handling of complex control flows. In addition, detection capability is maintained even as the number of functions increases (0.71 Accuracy for 13-14 functions), suggesting successful modeling of inter-function dependencies. Notably, the model maintains consistent performance even in the highest complexity bins across multiple metrics, with no significant degradation on complex samples. For instance, it achieves 0.75 accuracy on samples with high CC (196 mean). This stability across complexity metrics demonstrates the model's ability to handle realistic project-level codebases without being overwhelmed by increased complexity. 

\textcolor{black}{
\textbf{Performance Analysis on Post-Knowledge-Cutoff CVEs.} To evaluate our model's generalization capabilities on emerging vulnerabilities, we compiled a balanced dataset comprising 60 samples from CVEs published in 2025.
}
\textcolor{black}{
To construct this dataset, we crawled the NVD for CVEs published between January 1, 2025, and May 12, 2025, filtering for those associated with the following CWEs: CWE-119, CWE-120, CWE-121, CWE-122, CWE-125, CWE-190, CWE-415, CWE-416, and CWE-787. We focused exclusively on CVEs with public references linking to GitHub or GitLab commits. Where applicable, we also resolved repositories through known official mirrors. For instance, while the Linux kernel’s primary codebase is hosted at git.kernel.org, it has an official mirror at github.com/torvalds/linux.
}
\textcolor{black}{
In total, we crawled 1,583 CVEs. Of these, 1,194 did not include a Git commit in their public references. Among the remaining 389 CVEs, only 121 had valid CWE tags (i.e., not labeled as NVD-CWE-noinfo). Out of those, only 81 belonged to one of the CWEs listed in Table \ref{tab:cwe_distribution}, namely CWE-190, CWE-416, and CWE-125. We included CWE-125 despite our suboptimal performance on it due to limited training samples.
}
\textcolor{black}{
Among the 81 relevant CVEs, only 57 had commit files that fit within the input context of our model (32k tokens, with potential extension to 128k tokens given additional resources, see Sec. \ref{sec:discussion} for more details). Each of these 57 CVEs is paired with its corresponding patch, resulting in a total of 114 samples, balanced across the \textit{Vulnerable} and \textit{Safe} labels in our final dataset.
}

\textcolor{black}{
 On this dataset, \systemclassifier achieved an F1-score of 0.617 and Accuracy of 0.600, comparable to its performance on the ReposVul dataset. This demonstrates the model's robust generalization to novel vulnerability patterns that emerged after its knowledge cutoff, suggesting effective learning of fundamental vulnerability characteristics rather than mere memorization of known CVE instances.
}
\textcolor{black}{
Table \ref{tab:performance-on-reposvul_knowledge_cutoff} shows our performance on ReposVul and Post-Knowledge-Cutoff datasets.
}
\begin{table}[!htbp]
    \centering
    \caption{\textcolor{black}{Average Performance of \system on ReposVul \& 2025 Post-Knowledge-Cutoff (PKCO-25) datasets}}
    \label{tab:performance-on-reposvul_knowledge_cutoff}
    
    \resizebox{\columnwidth}{!}{
    \begin{tabular}{lcccc}
    \toprule
     Dataset & Accuracy & Precision & Recall & F1-score \\
     \midrule
     ReposVul &  0.634 & 0.542 & 0.700 & 0.610 \\ 
     PKCO-25 & \textcolor{black}{0.600} & \textcolor{black}{0.592} & \textcolor{black}{0.644} &  \textcolor{black}{0.617} \\ 
    \bottomrule
    \end{tabular}
    }
\end{table}

\textbf{Limitations.} During our analysis, we notice a performance degradation on few samples with very high nesting depth (>7 levels), dropping to 0.33 accuracy. While the model maintains reasonable performance across function count increases, the accuracy variance (0.33-0.84) suggests room for improvement in modeling some extremely complex inter-function relationships. These results represent a significant step toward practical vulnerability detection at the project level, though they also highlight specific areas where current approaches can be enhanced. \system's ability to maintain consistent performance across most complexity metrics, despite not being trained on project-level samples, demonstrates its potential for real-world deployments.

\subsection{Misclassification Analysis}
\label{subsec:eval_error_analysis}

This section presents a comprehensive error analysis of the \system's performance across multiple datasets, including FormAI, PrimeVul, ReposVul, and SVEN. 
In other words, we analyze errors made by \system~under the optimal threshold of each dataset to identify error sources, explain performance variations, and provide insights for future improvements. Table \ref{tab:performance_metrics} shows the performance of \system~by CWE.

\begin{table}[!htpb]
\caption{Performance Metrics of \system~by CWE Category}
\label{tab:performance_metrics}
\centering
\resizebox{\columnwidth}{!}{
\begin{tabular}{lllll}
\toprule
 & Accuracy & Precision & Recall & F1 Score  \\
\midrule
CWE-119 & 0.684 & 1.000 & 0.608 & 0.756  \\
CWE-120 & 0.333 & 1.000 & 0.111 & 0.200  \\
CWE-125 & 0.500 & 0.579 & 0.375 & 0.455  \\
CWE-190 & 0.582 & 0.681 & 0.688 & 0.684  \\
CWE-415 & 0.691 & 0.891 & 0.721 & 0.797  \\
CWE-416 & 0.618 & 0.762 & 0.557 & 0.643  \\
\bottomrule
\end{tabular}
}
\end{table}

\textbf{Performance on Trained Datasets.} \system~exhibit strong performance on several memory-related Common Weakness Enumeration (CWE) categories within the FormAI and PrimeVul datasets used for training. Specifically, the model achieves high accuracy and F1-scores for CWE-119 (Buffer Overflow), CWE-190 (Integer Overflow), CWE-415 (Double Free), and CWE-416 (Use After Free). This success can be attributed to the balanced representation of these CWEs in the training data and the model's ability to capture distinct code patterns associated with these vulnerability types.


\textbf{Challenges with Underrepresented CWEs.} Despite the model's strong performance on some memory-related CWEs, it faces challenges in detecting CWE-120 (Classic Buffer Overflow) and CWE-125 (Out-of-bounds Read). \khang{The low performance on these CWEs can be attributed to the limited number of examples, as shown in Table~\ref{tab:cwe_distribution}. Their underrepresentation in the training dataset likely contributed to the model's reduced effectiveness. Additionally, these vulnerabilities often involve subtle code variations and complex memory access patterns, which are inherently more difficult for the model to capture—especially under conditions of limited training data.}


\textbf{Generalization Performance on Unseen Datasets.} A critical aspect of evaluating the \system's effectiveness is its ability to generalize to unseen datasets. Despite the challenges posed by the ReposVul and SVEN datasets, which were not used during training, our model demonstrates promising generalization capabilities. The model achieved accuracy scores exceeding 0.6 on both datasets, indicating its ability to detect vulnerabilities in diverse, real-world codebases. While the model demonstrates promising generalization performance on unseen datasets, there is still room for improvement. We aim to bridge this gap by compiling a new high-quality dataset by synthetically augmenting real-world datasets spanning multiple files and functions and verifying the output with bounded model checking. 

\subsection{Robustness to Code Augmentation}
\label{subsec:eval_robusteness}

Robustness in vulnerability detection refers to a model's ability to maintain consistent performance when the input code undergoes semantically-preserving transformations. Robustness against code transformations indicates the model's ability to capture fundamental vulnerability patterns rather than superficial code characteristics. Robustness is also essential for security applications, as adversaries might attempt to evade detection by applying simple code transformations while maintaining the vulnerable behavior.

In this section, we study the ability of \system to handle different types of noise and variations. Risse et al. \cite{bohme_limits} define different data transformations to evaluate the performance of state-of-the-art vulnerability detection models and inspect their reliance on unrelated features. Risse’s results indicate that state-of-the-art models overfit unrelated features.

We select four code transformation algorithms, one from each category defined in Risse’s work \cite{bohme_limits}. The used transformations are described in Table \ref{tab:transformations}.

\begin{table}[h]
    \centering
    \caption{The semantic preserving transformations that we use in our experiments.}
    \label{tab:transformations}
    \resizebox{\columnwidth}{!}{%
    \begin{threeparttable}
    
    \begin{tabular}{lll}
    \toprule
    Identifier & Type & Description \\
    \midrule
    T1 & Identifier Renaming & Rename all function parameters to a random token. \\ 
    T2 & Statement Insertion & Insert unexecuted code. \\ 
    T3 & Statement Reordering & Move the code of the function into a separate function. \\ 
    T4 & Statement Removal & Remove all comments. \\
    \bottomrule
    \end{tabular}
    \end{threeparttable}
    }
\end{table}

\textbf{Experimental Setup.} To evaluate the robustness of \system, we conduct experiments using three diverse datasets: FormAI, PrimeVul, and SVEN. For each code slice in these datasets, we apply four previously mentioned code transformations. These transformations are designed to reflect common code modifications that preserve program semantics but may challenge model performance. Finally, we evaluate \system's performance on both the original and transformed code slices.


\begin{table}[!htpb]
\centering
\caption{Comprehensive Performance Results Across Datasets and Transformations}
\label{tab:robustness_results}
\fontsize{6.72}{9}\selectfont
\begin{tabular}{llcccc}
\hline
\textbf{Dataset} & \textbf{Transformation} & \textbf{Accuracy} & \textbf{Precision} & \textbf{Recall} & \textbf{F1-Score} \\
\hline
\multirow{4}{*}{FormAI}
& Normal and T4 & 0.8146 & 0.8097 & 0.8054 & 0.8075 \\
& T1 & 0.8146 & 0.8098 & 0.8054 & 0.8076 \\
& T2 & 0.8068 & 0.8000 & 0.8000 & 0.8000 \\
& T3 & 0.8355 & 0.8506 & 0.8000 & 0.8245 \\
\hline
\multirow{4}{*}{PrimeVul}
& Normal and T4 & 0.7250 & 1.0000 & 0.4500 & 0.6206 \\
& T1 & 0.7375 & 1.0000 & 0.4750 & 0.6441 \\
& T2 & 0.6650 & 1.000 & 0.3250 & 0.4906 \\
& T3 & 0.6750 & 0.6750 & 0.6750 & 0.6750 \\
\hline
\multirow{4}{*}{SVEN}
& Normal and T4 & 0.6020 & 0.5590 & 0.9534 & 0.7048 \\
& T1 & 0.5551 & 0.5488 & 0.6977 & 0.6143 \\
& T2 & 0.6220 & 0.6279 & 0.6279 & 0.6279 \\
& T3 & 0.6220 & 0.5864 & 0.8682 & 0.7000 \\
\hline
\end{tabular}
\end{table}

Table \ref{tab:robustness_results} shows the results of this experiment, which reveal several interesting patterns:

\begin{enumerate}
    \item \textbf{Complete robustness against comments removal}: \system is unaffected by the comment removal transformation (T4) because our slice construction approach inherently ignores comments, focusing solely on execution paths. As a result, the constructed slices are free of comments by design.
    
	\item \textbf{Dataset-Dependent Robustness}: The model shows varying levels of robustness across different datasets. Notably, it demonstrates the highest robustness on the FormAI dataset, where performance improves slightly under transformations (F1-score increase of 2.10\%).
    
	\item \textbf{Transformation Impact}: Among the transformations, T3 (function extraction) generally had the most significant impact on the model's performance, particularly affecting recall. This suggests that the slice-based approach is somewhat sensitive to changes in function boundaries.
    
	\item \textbf{Precision-Recall Trade-off}: For PrimeVul, we observe an interesting trade-off where transformations lead to decreased precision but improved recall, indicating that the model becomes more conservative in its vulnerability predictions under code transformations.
\end{enumerate}

The robust performance of \system, particularly on the FormAI dataset, can be attributed to two key factors. First, our CPG-based slice construction inherently focuses on semantic relationships rather than syntactic features, making it naturally resistant to surface-level code changes. Second, by constructing slices that capture essential vulnerability-related interactions, we effectively filter out irrelevant code modifications.

These findings highlight the importance of semantic-aware vulnerability detection approaches and suggest that future improvements should focus on maintaining robust performance across diverse vulnerability patterns while preserving the ability to capture essential semantic relationships in the code.




\section{Discussion}
\label{sec:discussion}

\textbf{Limitations in Vulnerability Type Coverage.} Current code property graph (CPG) approaches, while effective for many vulnerability types, face inherent limitations in modeling certain classes of security flaws. As demonstrated by Yamaguchi et al. \cite{joern}, vulnerabilities like race conditions and design errors remain challenging to express using graph traversals since they often depend on runtime properties or require deeper understanding of the system's intended design. This limitation stems from the static nature of CPG analysis, which cannot capture dynamic program behaviors or complex architectural decisions. 

\noindent \textbf{Dataset Quality and Availability.} A significant challenge in vulnerability detection research is the scarcity of high-quality datasets. As evidenced in the PrimeVul study, existing benchmarks suffer from poor data quality, low label accuracy, and high duplication rates. For instance, their analysis revealed that only 38\%-64\% of functions labeled as vulnerable in popular datasets actually contained security flaws. The ReposVul paper further highlights this issue by demonstrating how vulnerability-fixing commits often include unrelated code changes, leading to noisy labels when using automated collection methods. This data quality problem fundamentally limits the effectiveness of machine learning approaches for vulnerability detection.

\textbf{Project-level Vulnerability Detection.} The promising performance of \system on project-level vulnerability detection, despite being primarily trained on function-level data, suggests significant potential for advancing automated security analysis of complex software systems. While achieving 0.634 accuracy on ReposVul demonstrates meaningful progress, the pipeline's performance variance across different complexity metrics indicates opportunities for further enhancement. A critical path forward lies in the development of high-quality, project-level vulnerability datasets that capture the intricate dependencies and interactions present in real-world codebases. 
The creation of these curated datasets, combined with \system's  capability to handle complex code structures through its CPG-guided approach, could significantly advance the state-of-the-art in project-wide vulnerability detection. This represents a promising direction for future research, potentially enabling more comprehensive and reliable security analysis of large-scale software projects.

\textcolor{black}{
\textbf{Limitations of Binary Classification.} While \system does output a binary classification (vulnerable/safe), we designed our approach to mitigate reasoning opacity through several mechanisms. The CPG-based slice construction inherently preserves reasoning pathways by capturing execution flows and data dependencies that contribute to vulnerability presence. For example Figure \ref{fig:slice_extraction} provides visibility into how specific code elements interact to create vulnerable conditions. In order to explore the reasoning effect on this task, we fine-tuned \systemclassifier with reasoning traces extracted from DeepSeek-v3, but the results did not improve as expected as the accuracy was similar to binary classification. This was likely due to the quality and quantity of available reasoning traces. Recent distillation work \cite{guo2025deepseek} (e.g., from DeepSeek-R1 to Qwen-2.5) demonstrates that effective reasoning transfer requires substantial data volumes (i.e., 800k datapoints), which exceeded our available data for this specific task. Enhancing vulnerability reasoning through high-quality traces remains a future work.
}

\textcolor{black}{\textbf{Context Length Limitation of the Base LLM for Query Generation.}
While the base model for LLMxCPG-Q (Qwen/Qwen2.5-Coder-32B-Instruct) supports a 128K token context, we fine-tuned it with a 32K token limit due to constraints in our available computing power. It is worth noting that while versions of Qwen (and other models) with even larger context windows (e.g., 1M tokens) are available, fine-tuning these necessitates substantial computational resources, which were beyond our current capacity. Consequently, the query generation process with LLMxCPG-Q is most effective for codebases or commit files that fit within this 32K fine-tuned context length (and theoretically 1M given the necessary resources).
}
\textcolor{black}{
Importantly, this context length limitation was specific to the query model (LLMxCPG-Q) and did not pose a problem for our detection model (LLMxCPG-D, based on Qwen/QwQ-32B-Preview). The Code Property Graph (CPG) based slicing was indeed effective in reducing the code to a manageable size for the detection model (even when limting the context to 8K in LLMxCPG-D for faster generation), ensuring its input context was not exceeded. This addresses the core function of CPG in our framework – to precisely slice and reduce code to fit the detection model's context effectively.
}


\section{Related Work}
\label{sec:related_work}

\textbf{Deep Learning Approaches.}
The evolution of deep learning in vulnerability detection progresses through several architectural paradigms. Initial approaches centered on sequential modeling, with VulDeePecker \cite{li2018vuldeepecker} establishing the viability of LSTM networks for processing code gadgets. SySeVR \cite{li2021sysevr} advanced this foundation by introducing systematic feature extraction based on semantic relationships. A pivotal investigation by Chakraborty et al. \cite{chakraborty2021deep} revealed a critical limitation: while deep learning models demonstrated promising results, their decision-making often relied on superficial code patterns rather than fundamental vulnerability characteristics.
Architectural innovations emerged to address these limitations. LineVul \cite{fu2022linevul} introduced transformer-based architectures for fine-grained vulnerability detection at the line level, while Steenhoek et al. \cite{steenhoek2023dataflow} incorporated dataflow analysis principles into deep learning frameworks to enhance detection efficiency. Graph-based representations have proven particularly effective, with several notable implementations. Vul-LMGNN \cite{liu2024source} achieved superior results by integrating pre-trained code language models with code property graphs through a specialized gated Graph Neural Network architecture. FUNDED \cite{wang2020combining} enhanced the reliability of graph-based approaches through automated data acquisition and probabilistic learning mechanisms. ReGVD \cite{nguyen2022regvd} further refined graph neural network architectures, demonstrating substantial performance improvements through targeted architectural modifications.

\noindent \textbf{LLMs for Vulnerability Detection.}
Large Language Models (LLMs) have emerged as a transformative approach to vulnerability detection, though recent research has revealed important nuances in their application. PrimeVul \cite{primevul} provided critical insights by demonstrating that conventional benchmarks substantially overestimate model performance, necessitating more rigorous evaluation methodologies. VulBERTa \cite{hanif2022vulberta} established that domain-specific pre-training protocols significantly enhance detection capabilities, while VulSim \cite{shimmi2024vulsim} introduced an innovative multi-dimensional embedding approach that simultaneously captures semantic, contextual, and syntactic code properties.
Recent advances have focused on specialized fine-tuning strategies and architectural integration. VulLLM \cite{du2024generalization} developed a multi-task instruction fine-tuning framework that demonstrably improves model generalization across diverse vulnerability types. MSIVD \cite{yang2024security} advanced this direction through carefully designed instruction sets and decomposed task structures, achieving enhanced detection accuracy. Hybrid architectures have shown particular promise, with VDDA \cite{chang2023vdda} successfully combining deep learning with attention mechanisms, and CPVD \cite{zhang2023cpvd} enabling cross-project vulnerability detection through graph attention networks. Foundational models including UnixCoder \cite{guo2022unixcoder}, CodeBERT \cite{feng2020codebert}, and GraphCodeBERT \cite{guo2021graphcodebert} have demonstrated the value of incorporating structural code information during the pre-training phase, establishing essential building blocks for future advances in the field.


\system, differs from previous approaches by uniquely leveraging LLMs to generate valid CPGQL queries for traversing code property graphs. Unlike traditional deep learning approaches that directly learn from code representations, or pure LLM approaches that may miss structural information, \system uses LLMs to guide the graph traversal process itself. While VulLLM and MSIVD demonstrate the potential of fine-tuned LLMs, and Vul-LMGNN shows promise in combining LMs with graph neural networks, our approach maintains interpretability through explicit query generation. This novel integration preserves the benefits of graph-based program analysis while utilizing LLMs' pattern recognition capabilities in a more controlled and explainable manner.

\section{Conclusion}
\label{sec:conclusion}

In this paper, we have presented \system, a novel vulnerability detection approach that effectively addresses fundamental limitations in current deep learning-based methods. By combining Code Property Graphs with Large Language Models, our framework achieves superior performance across multiple evaluation dimensions. The empirical results demonstrate substantial improvements over existing approaches, with F1-score increases of up to 40\% and consistent performance on rigorously verified datasets. \system's ability to maintain robust detection capabilities under code transformations while generalizing effectively to complex, multi-function codebases represents a significant advancement in automated vulnerability detection. These results establish \system as a promising foundation for future research in software security analysis.

\section*{Acknowledgments}

\textcolor{black}{
We thank the anonymous reviewers for their valuable feedback and their help to improve the quality of this manuscript.
}

\section*{Ethics Considerations}

Our vulnerability detection research adheres to responsible disclosure protocols and established security research guidelines. We carefully balance the benefits of identifying security weaknesses against potential risks. All discovered vulnerabilities are reported through appropriate channels, allowing sufficient time for patches before public disclosure. We maintain strict confidentiality throughout the research process and ensure our methods do not compromise system integrity or user privacy.

\section*{Open Science}

Our source code, fine-tuned models, and testing datasets is
available publicly to the community to foster research in
this field at: \url{https://github.com/qcri/llmxcpg} and \url{https://zenodo.org/records/15614095}.

\bibliographystyle{plain}

\appendix




\section{Code Metrics}
\label{apdx:code_metrics}

This section defines foundational software metrics utilized in quantitative code analysis, with particular emphasis on structural and cognitive complexity assessment.

\textbf{Lines of Code (LOC).} A fundamental volumetric metric quantifying program size through source code line enumeration. LOC encompasses physical lines containing executable statements, declarations, and definitions, while excluding blank lines and comments. This metric serves as a primary indicator of implementation scale and maintenance burden.

\textbf{Cyclomatic Complexity (CC).} A graph-theoretic metric measuring program flow complexity through control flow analysis. CC quantifies the number of linearly independent paths through program source code, calculated as:
\begin{equation}
CC = E - N + 2P
\end{equation}
where:
\begin{itemize}
\item $E$ represents the number of edges in the control flow graph
\item $N$ represents the number of nodes
\item $P$ represents the number of connected components
\end{itemize}

For a given function $f$, the complexity can be alternatively expressed as:
\begin{equation}
CC(f) = 1 + \sum_{d \in D} p(d)
\end{equation}
where $D$ is the set of decision points and $p(d)$ represents predicates at each decision point.

\textbf{Number of Functions.} A modularity metric quantifying discrete functional units within the codebase. This metric encompasses all function declarations and definitions, including methods, procedures, and subroutines, providing insight into code compartmentalization and potential maintenance complexity.

\textbf{Number of Branches.}
A control flow metric enumerating decision points within the code. This encompasses conditional statements (if-else constructs), switch cases, and loop conditions. The total branch count $B$ for a program $P$ can be expressed as:
\begin{equation}
B(P) = \sum_{i=1}^{n} b_i
\end{equation}
where $b_i$ represents individual branching constructs.

\textbf{Nesting Depth.}
A structural complexity metric measuring the maximum level of control structure embedding within the codebase. For a given code block $c$, the nesting depth $ND$ is defined as:
\begin{equation}
ND(c) = \max_{s \in S} d(s)
\end{equation}
where $S$ represents the set of all statements in the code block and $d(s)$ represents the nesting level of statement $s$.

\section{Comparison with Other Base Models}
\label{apdx:comparison_with_base_models}

\textcolor{black}{
In order to choose the final model that we used in final detection (i.e., \systemclassifier), we fine-tuned Phi-4 (14B), Qwen2.5-Coder (32B), Codestral (22B), and QwQ-Preview (32B). Table \ref{tab:model-comparison} shows the comparison among the models on PrimeVul dataset.
}
\textcolor{black}{
\begin{table}[!htbp]
\caption{Comparison of fine-tuned models on the PrimeVul dataset.}
\label{tab:model-comparison}
\centering
\begin{tabular}{lcc}
\toprule
\textbf{Model} & \textbf{Accuracy} & \textbf{F1-score} \\
\midrule
Phi-4 (14B) & 0.5912 & 0.5356 \\
Codestral (22B) & 0.6233 & 0.5521 \\
Qwen2.5-Coder (32B) & 0.6823 & 0.6001 \\
QwQ-Preview (32B) & 0.7250 & 0.6206 \\
\bottomrule
\end{tabular}
\end{table}
}

\textcolor{black}{
\section{Impact of Slice Construction and Query Generation Model}
\label{apdx:impact_of_slice_construction_and_query_generation_model}
To assess the impact of code slicing on detection performance, we evaluate \systemclassifier on the test datasets using the full code, bypassing the CPG slicing step. The model's performance consistently declines across all datasets when slicing is omitted: FormAI achieves an accuracy of 0.6762 (down from 0.8146 with slicing), PrimeVul drops to 0.4875 (from 0.7250), and SVEN falls to 0.5078 (compared to 0.6020).
}

\textcolor{black}{
To further demonstrate the flexibility of the query-based model, we employ Joern-scan\footnote{\url{https://docs.joern.io/scan/}}, a tool that executes predefined queries targeting various vulnerability patterns, to analyze the 50 samples previously selected for the semantic correctness experiment (see Section \ref{subsubsec:eval_query_generation}). Notably, Joern-scan fails to detect any of the vulnerable samples in this subset, as it relies on a fixed set of sensitive function calls in C. In real-world scenarios, however, developers often implement custom wrappers around these functions, making them more difficult to detect using static query-based approaches.
}

\section{Choosing a Threshold}
\label{apdx:choose_threshold}

\begin{figure}[!htbp]
    \centering
    \includegraphics[width=3.3in]{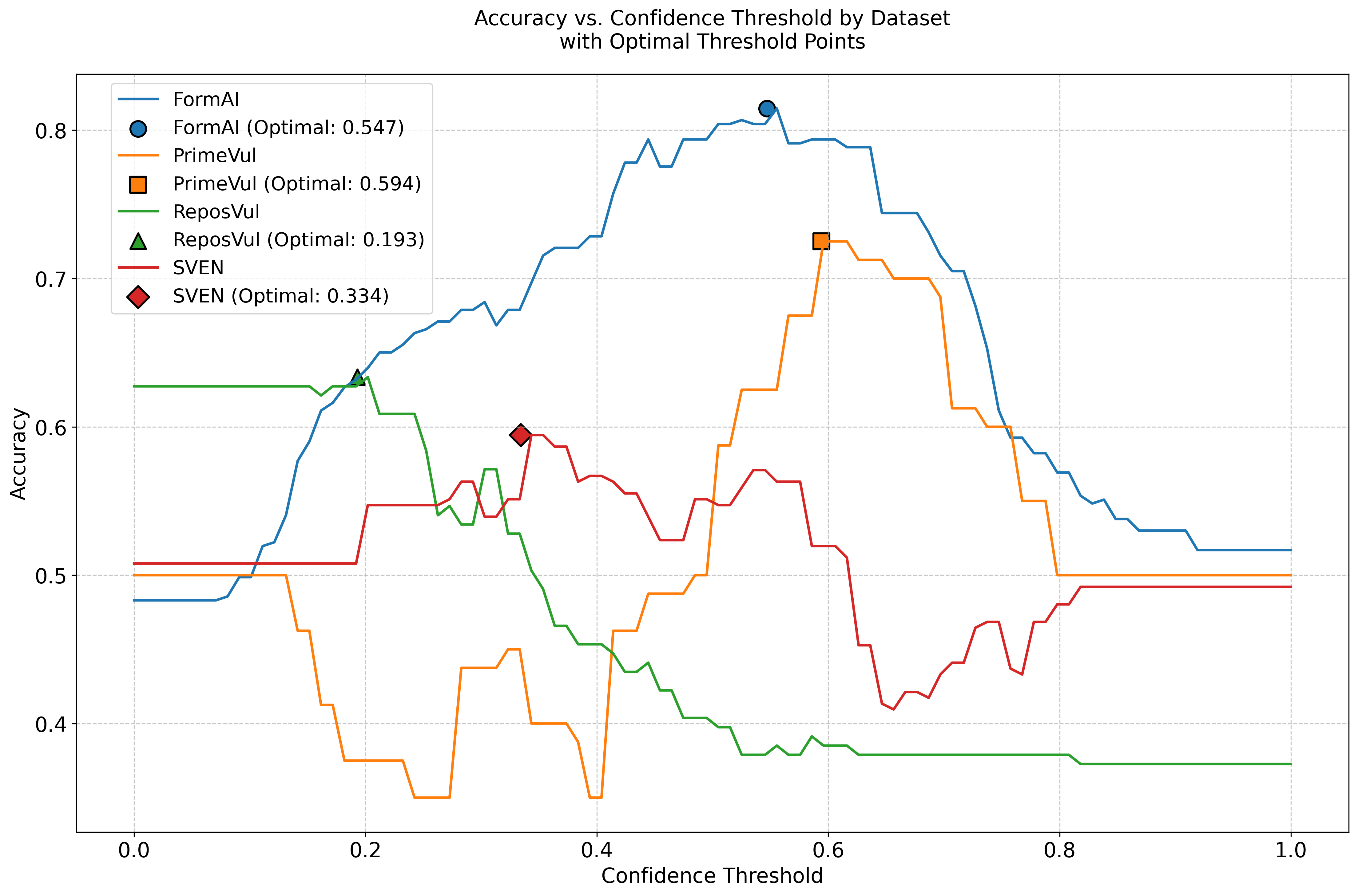}
    \caption{\systemclassifier's Accuracy vs Confidence Threshold for Different Datasets.}
    \label{fig:threshold_accuracy_analysis}
\end{figure}

The process of choosing a threshold for the model starts by select few labeled datapoints (e.g., 20 was used for our case) from the validation splits of the target datasets. Then, we generate predictions with \systemclassifier and maximize the accuracy by performing a complete search over the interval of threshold values $[0, 1]$. Figure \ref{fig:threshold_accuracy_analysis} shows the effect of vaying threshold on the accuracy of  \systemclassifier on different datasets.

\section{Prompt Templates}
\label{apdx:promt_templates}

In this section, we present the employed prompts for query generation and vulnerability detection.

\begin{figure}[!htpb]
\centering
\small
\begin{tcolorbox}[
    colback=gray!5!white,
    colframe=black,
    sharp corners,
    boxrule=0.75pt
    ]
    \textbf{Instruction:}\\
        \textit{Your task is to design Precise Joern CPGQL Queries for Vulnerability Analysis.}
        \\
        \\
        \textbf{Objective:}\\
        Develop targeted CPGQL Joern queries to:
        \begin{itemize}
            \item Identify taint flows based on your analysis.
            \item Capture potential vulnerability paths.
        \end{itemize}

        \textbf{Constraints:}
        \begin{itemize}
            \item Queries must be executable in Joern/CPGQL
            \item Use Scala language features for query construction
            \item Last query must use reachableByFlows to identify vulnerable paths
        \end{itemize}

        \textbf{Output Requirements:}\\
        Provide a JSON object with one field "queries": Sequence of CPGQL queries to detect vulnerability

        \vspace{0.5em}

        \textbf{Expected JSON Output Format:}\\
        \{
        
        \hspace{1em} "queries": ["Query1"\hspace{1em}, "Query2", ..., "Final \\
        \text{\hspace{1em}} Reachable Flows Query"]
          
        \}

        \vspace{0.5em}

        \textbf{Example Output:}\\
        Example in Figure \ref{fig:example_queries}

        \vspace{2em}

    \textbf{Input:}
    <Code>
\end{tcolorbox}

\caption{Prompt to generate CPGQL queries.}
\label{fig:query_gen_prompt}
\end{figure}

\begin{figure}[h]
    \centering
    \includegraphics[width=3.3in]{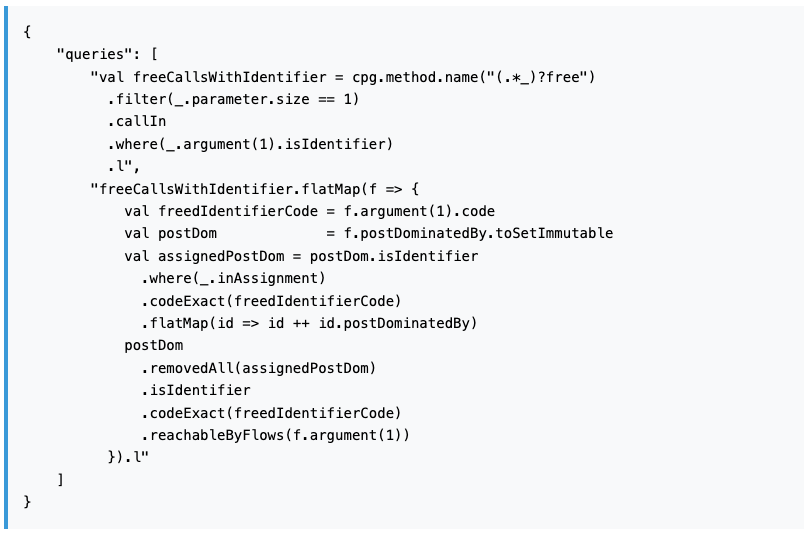}
    \caption{Example CPGQL Queries.}
    \label{fig:example_queries}
\end{figure}

\begin{figure}[!htpb]
\centering
\small
\begin{tcolorbox}[
    colback=gray!5!white,
    colframe=black,
    sharp corners,
    boxrule=0.75pt
    ]
    \textbf{Instruction:}\\
    \textit{You are a security code vulnerability analyzer. Your task is to carefully analyze the provided code snippet. Note that the provided code snippet might not be complete, but it has all the important context.
    }
    \vspace{0.5em}
    
    Your output must be EXACTLY ONE WORD:
    \begin{itemize}
        \item If you detect any potential security vulnerability in the specified code segment, return: VULNERABLE
        \item If the code segment appears to be secure and free from obvious vulnerabilities, return: BENIGN
    \end{itemize}
    
    \vspace{0.5em}
    
    \textbf{IMPORTANT GUIDELINES:}\\
    Consider common vulnerability types such as:
    \begin{itemize}
        \item Buffer overflows
        \item Improper input validation
        \item Integer Overflow
        \item Memory corruption potential
        \item Double free
        \item Use after free
    \end{itemize}

    \vspace{0.5em}
    
    Your response must be either 'VULNERABLE' or 'SAFE' - no additional explanation

    \vspace{0.5em}
    
    \textbf{Output format:}\\
    One word: VULNERABLE or SAFE

    \vspace{1em}
    \textbf{Input:}
    <Code>
\end{tcolorbox}
\caption{Prompt to classify code slices.}
\label{fig:classifier_prompt}
\end{figure} 

\end{document}